\documentclass{icrc} 
 
\input psfig.sty 
 
\usepackage{times} 
\usepackage{graphicx} 

\begin{document} 
 
\title{Cosmic ray propagation and acceleration} 
 
\author[1]{Micha{\l} Ostrowski} 
 
\affil[1]{Obserwatorium Astronomiczne, Uniwersytet Jagiello\'nski, ul. 
Orla 171, 30-244 Krak\'ow, Poland} 
 
\correspondence{E-mail: mio\@@oa.uj.edu.pl} 
 
\firstpage{1} 
\pubyear{2001} 
 
 
\maketitle 
 
\begin{abstract} 
 
This report covers the studies of energetic particle propagation and 
acceleration presented during the 27th ICRC in sessions OG 1.3 and 1.4~, plus 
a few selected papers from neighbouring OG sections. We list problems 
discussed within these sections and we shortly describe selected 
interesting achievements and/or novel approaches. 
\end{abstract} 
 
\section{Introduction} 
 
In the present rapporteur's opinion, which is not always shared by 
other cosmic ray physicists, our knowledge of cosmic ray propagation and 
acceleration is only superficial and fragmentary now. Essentially, even 
the identification of basic acceleration processes is unsure, as well as 
understanding of details of these processes acting in shocks and other 
supersonic flows, or in turbulent media. Very nice fits of the proposed 
simple models to the observational data are quite often possible, but it 
is natural noting the fact that in many cases only single 
observational parameters are available, like the particle spectrum 
slope or the mean age of unstable nuclei. Therefore, to reach 
further progress in this field an era of studying complicated and/or 
elaborated models must begin, supported by substantial effort to 
increase the measurements' data base. Perhaps this personal attitude can 
explain my choice of problems to be discussed in this report.

Below, results of approximately 90 papers are reported. To make a 
reference to a given paper I will use either the more traditional form 
with authors names plus the page number in the proceedings (for clarity 
a symbol `p.' is given before the page number), or only the name HE, OG 
or SH, followed with the session number and the page number. In the 
first part of the present report, in section 2 presentations listed as 
the ones dealing with particle propagation are discussed (all OG 1.3 
papers + OG 1.2-p.1804, p.1812), while in the second part (Section 3) 
the acceleration papers (all OG 1.4 papers + OG 2.3-p.2701, p.2713 and 
2.4-p.2742) are reviewed. Then, a few final remarks are presented in the 
last section 4.

\section{Cosmic ray propagation} 
 
Precise measurements of cosmic ray abundances and spectra by present day 
experiments like CRIS, SCUBA, ACE, KASKADE, in various energy ranges, 
allow much more detailed modelling of cosmic ray transport within our 
galaxy's magnetic field structure and inhomogeneous gas distribution. 
Therefore, we start with a short reminder of the main observational 
facts related to these structures. Later we summarize results on cosmic 
ray propagation, divided into parts (subsections) describing various 
aspects. Selection of papers described in a given subsection is a 
somewhat arbitrary choice of the rapporteur. Let us start with pointing 
out a few serious difficulties for any attempt to construct a realistic 
model of cosmic ray transport within our galaxy and on larger scales.

\subsection{On the structure of the cosmic magnetic field, the matter 
distribution and the velocity field in our Galaxy} 
 
All studies of cosmic ray interactions face the fact of an extremely 
inhomogeneous matter distribution in the galactic disk (cf. Ferri\'ere 
2001). In our galaxy more than 50\% of diffuse matter is confined to 
discrete (molecular and cold neutral) clouds within $\sim 1 - 2$\% of 
the interstellar volume. These clouds are distributed close ($< 100 {\rm 
pc}$) to the galactic plane, non-uniformly both in radial and azimuthal 
directions, where the molecular torus for $3.5 < r < 7$ kpc and the 
spiral arm concentrations, dominate respectively. With respect to those 
cold and dense clouds, the density of the gaseous disk is 2 or 3 
orders of magnitude lower, and additional 2 orders of magnitude lower 
densities are considered for the galactic halo. At the same time the 
gaseous disk and halo have more than one or, respectively, two orders of 
magnitude larger volumes. 
 
To model the cosmic ray propagation inside the Galaxy and in the 
intergalactic space one needs sufficient information about the magnetic 
field structure, both the smooth large-scale component and the turbulent 
component responsible for particle scattering. A large ratio of the 
measured cosmic ray life time and the galaxy crossing time allows us to 
assume that the observed particle trajectories average over the galactic 
(+ halo) volume and the simple approaches based on the diffusive 
equation with the {\it effective} vertical diffusion coefficient can 
properly reproduce the mean parameters (age, abundances) of cosmic ray 
transport at low energies. However, if one intends for example to 
reproduce a real spatial structure of the cosmic ray density or of the 
mean age or grammage of particles, more detailed information about 
particle trajectories may be required. Unfortunately, there is far from 
full understanding of the magnetic field structure in space. Numerous 
studies show quite complicated patterns, most often observed as a 
projection on the celestial sphere. The observed magnetic field 
structures like `superbubbles' have vertical scales comparable to the 
height of the galactic gaseous disk, or even extend farther away if 
pulled out with the forming wind (cf. OG 1.3-1924). Thus a complicated 
magnetic connectivity may be formed within the galactic volume leading 
to substantial (?) variations of local properties of the cosmic ray 
population. 
 
At present information about the intergalactic magnetic field is only 
fragmentary (cf. Kronberg 2001), weakly constraining the propagation 
models of extremely high energy cosmic rays. Both large scale relatively 
uniform and strong ($\sim 1$ $\mu$G) supergalactic structures are 
considered as well as structures formed with the separate magnetized 
plasma blobs ejected from active galaxies during their cosmological 
evolution.

\subsection{Cosmic ray abundances and age} 
 
A substantial amount of data accumulated on cosmic ray primary and 
secondary abundances allows modelling different aspects of particle 
propagation. The papers we consider in this section present 
interpretation of LiBeB abundances and ages (OG 1.3-p.1831, p.1835, 
p.1938, p.1840, p.1937), role of K-capture processes in cosmic ray 
propagation (OG 1.3-p.1844), antiproton component (OG 1.3-p.1864, 
p.1868, p.1873, p. 1877) and anti-particle related problems (OG 
1.3-p.1880, p.1872), radioactive secondaries in cosmic rays (OG 
1.3-p.1860, p.1836), some other aspects of cosmic ray propagation (OG 
1.3-p.1892, p.1896). 
 
A number of presentations use simple propagation models to reproduce 
available measurements of cosmic ray abundances and ages. In some 
studies the cosmic ray data serve to impose limits on the model 
parameters, including important astrophysical issues of existence or not 
of the galactic wind and/or the role of particle re-acceleration in the 
interstellar medium. Unfortunately, besides the above discussed 
complicated astrophysical scenario, which may be improperly described by 
simple leaky-box or diffusive (+wind) models, the present knowledge of 
interaction cross sections is frequently insufficient to allow for 
definitive conclusions. In most cases authors were able to reproduce the 
magnitude of the measured particle fluxes, but the energy dependence 
suggested by the model was in some cases different compared to the 
one seen in the measurements. Nevertheless, let us note an interesting 
example of such a study. 
 
The radioactive electron capture isotopes can be used to 
investigate interstellar propagation of low energy cosmic rays due 
to the strong energy dependence of the surviving fractions in the 
interstellar medium. An interesting study intended to constrain particle 
reacceleration within the diffusive disk-halo mo\-del was presented by 
Jones et al. (p.1844), who considered the abundance modification 
of Cr, V and Ti isotopes by K-capture processes $^{51}{\rm Cr}$ $\to$ 
$^{51}{\rm V}$ and $^{49}{\rm V}$ $\to$ $^{49}{\rm Ti}$. Their results 
summarized in figure 1 show the importance of K-capture processes for such 
propagation study, but do not yield a solid conclusion on the role of 
the reacceleration process. The authors conclude that the statistical 
accuracy of the ACE experiment itself is high enough to see a 
signature of reacceleration, however, the uncertainties in nuclear 
cross-sections are probably too large to conclude that a reacceleration 
process took place. 
 
\begin{figure}                        
\vspace{67mm} 
\includegraphics{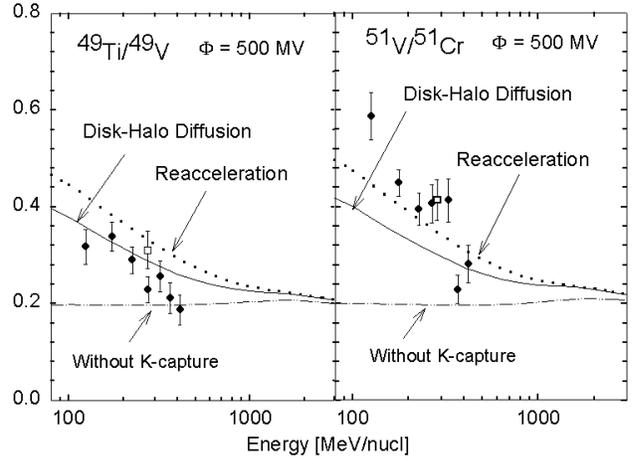} 
\caption{Ratios of $^{49}{\rm Ti / ^{49}{\rm V}}$ and 
$^{51}{\rm V} / ^{51}{\rm Cr} $: the ACE measurements compared to 
models (from Jones et al. p.1844).} 
\end{figure}

\begin{figure*}                           
\vspace{21.0cm} 
\includegraphics{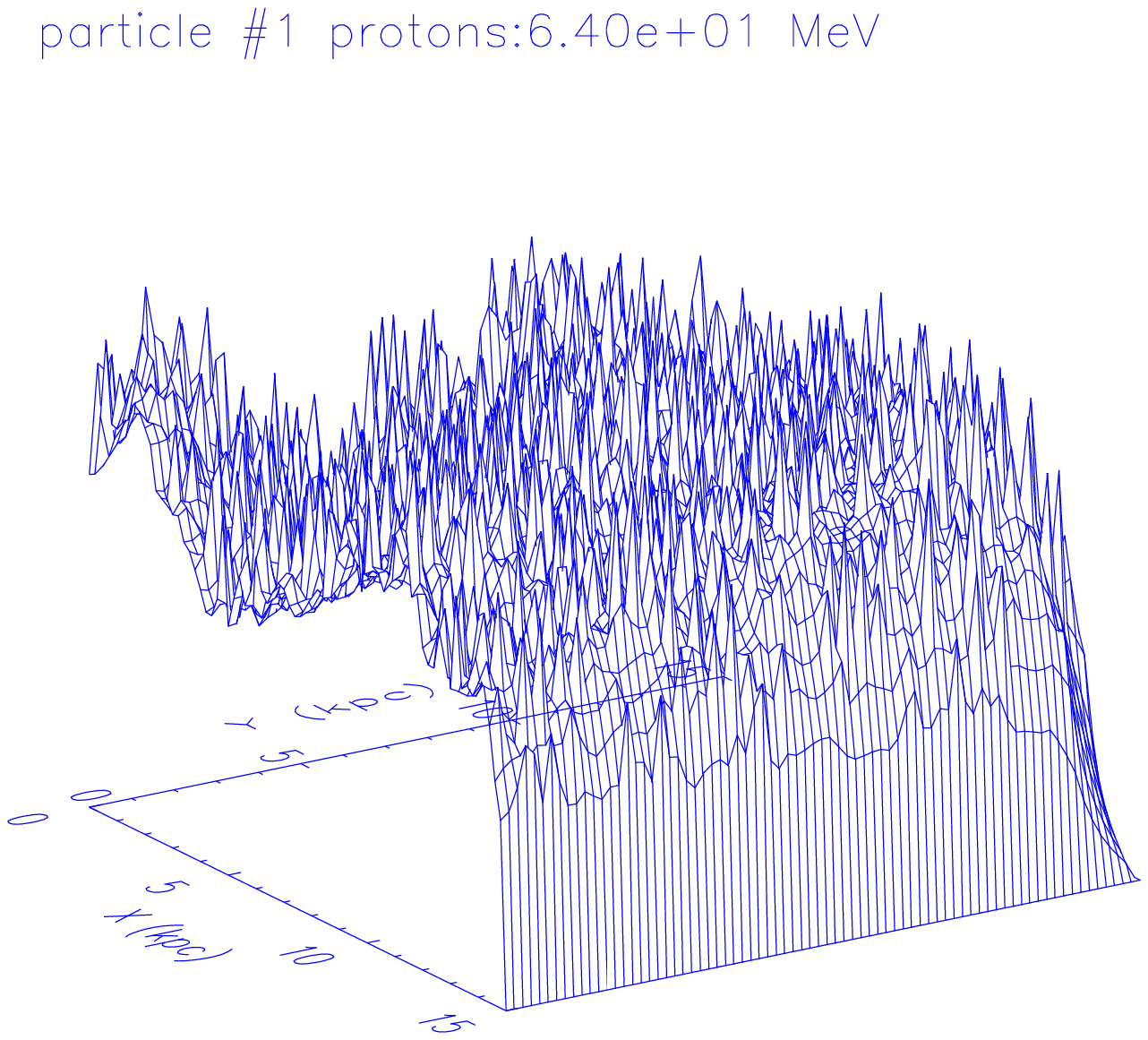} 
\includegraphics{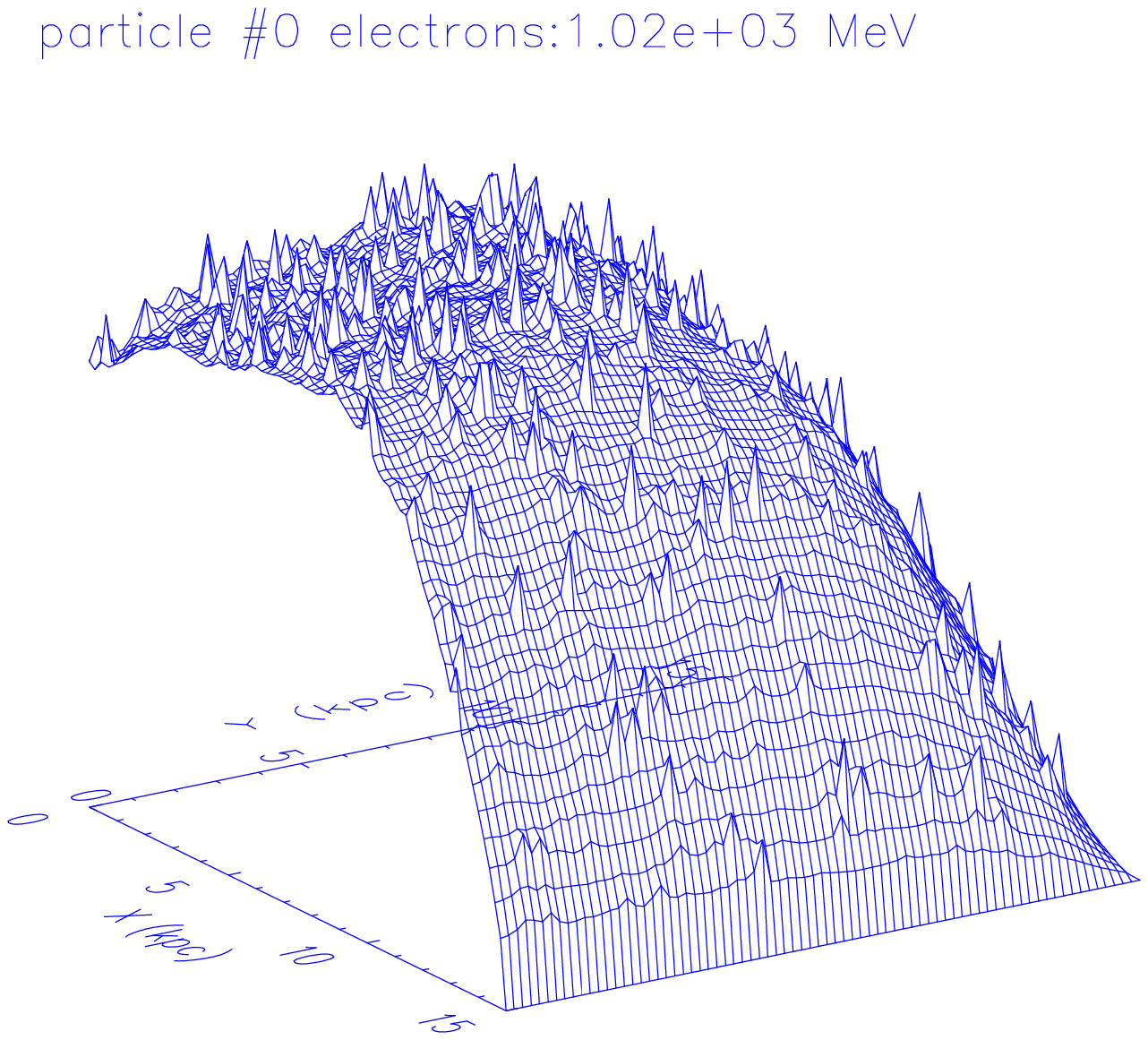} 
\includegraphics{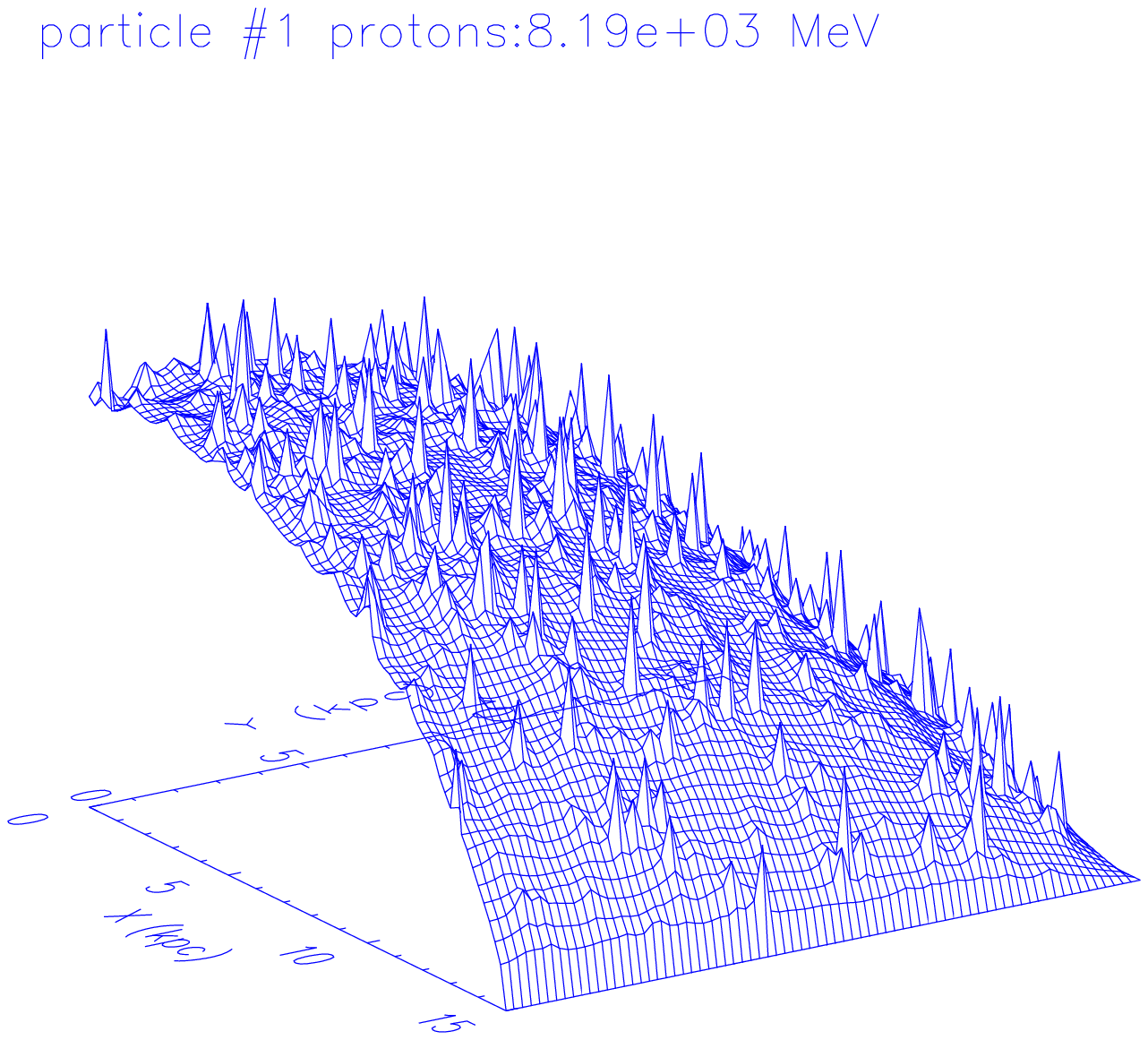} 
\includegraphics{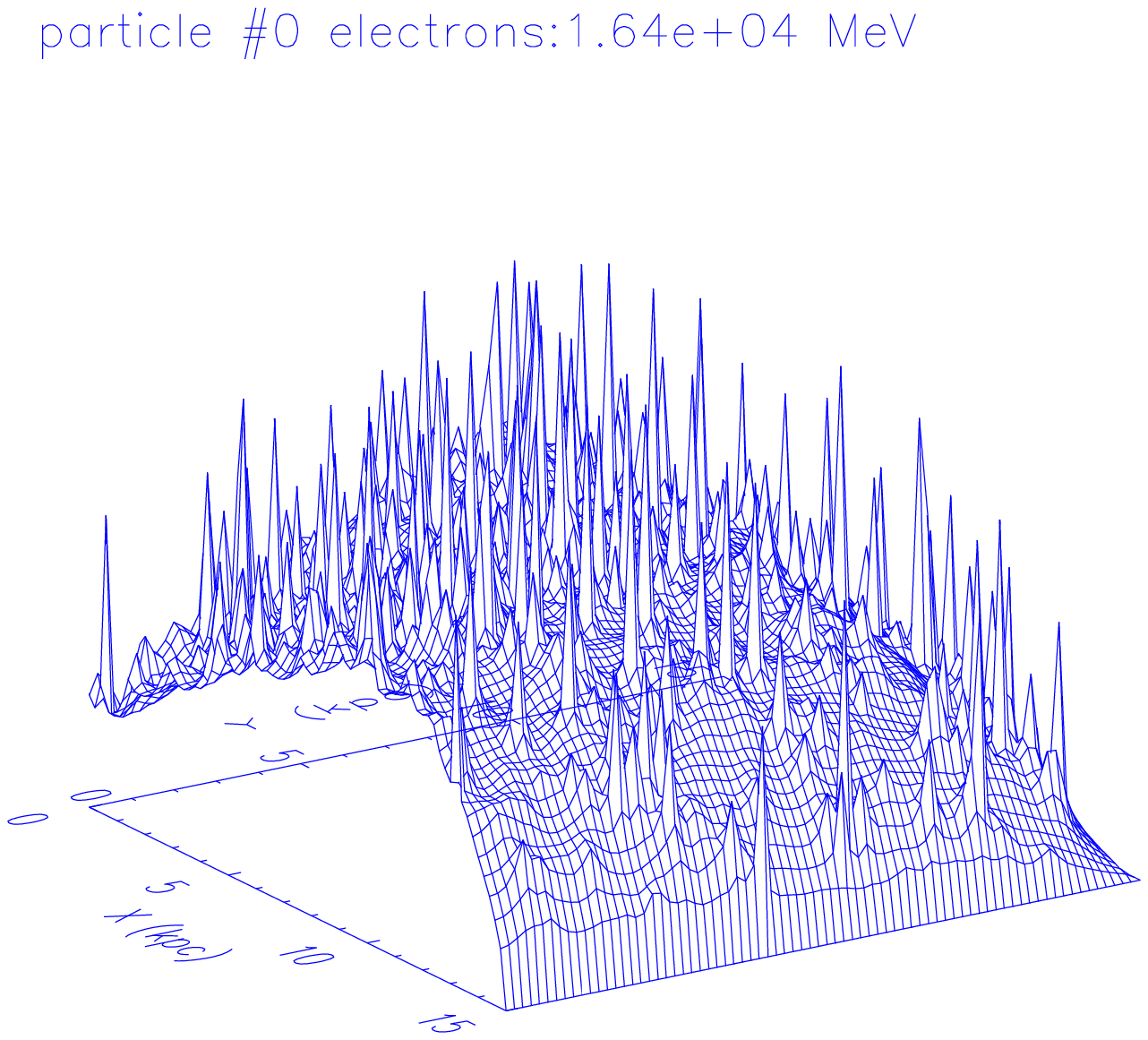} 
\includegraphics{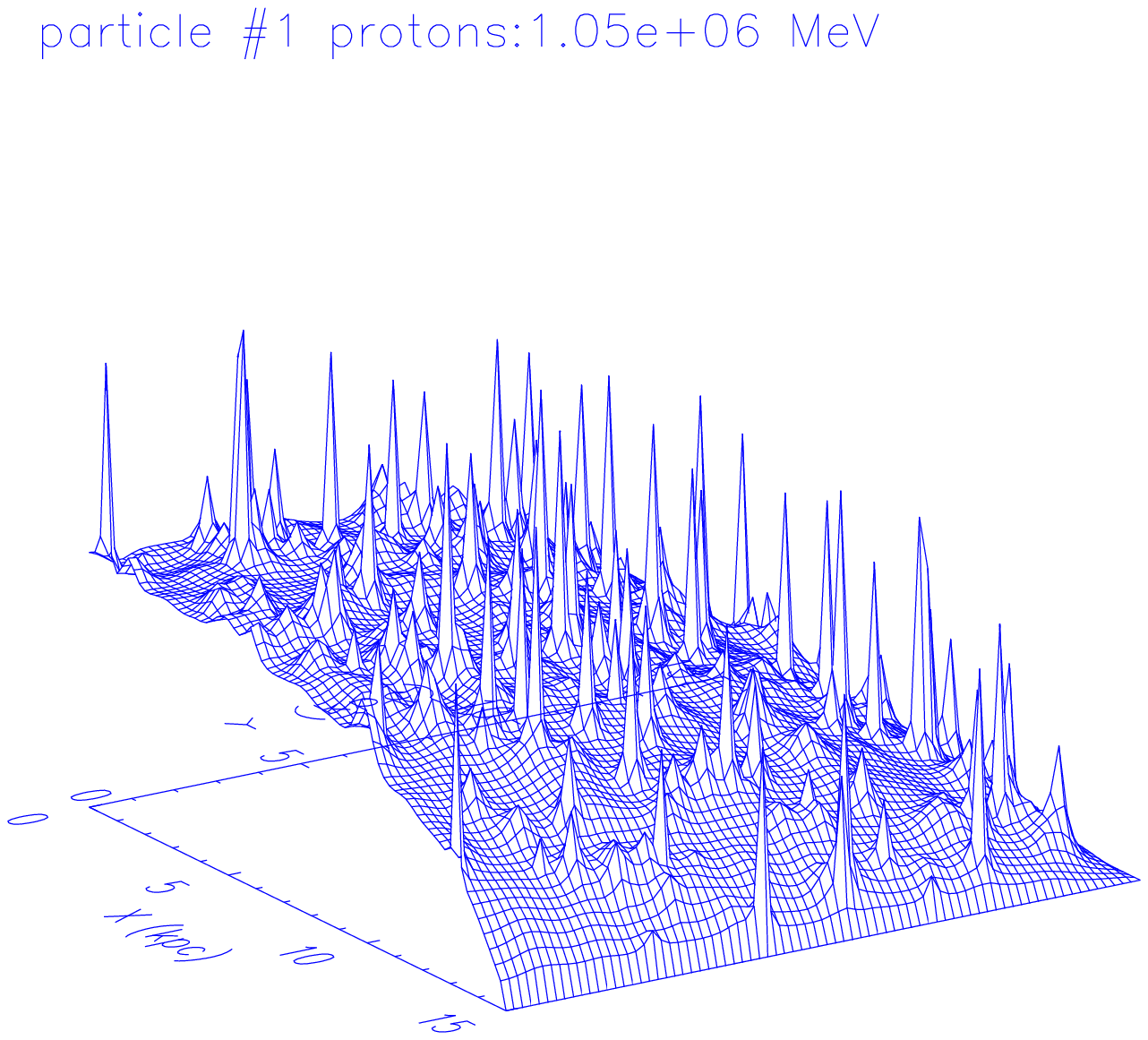} 
\includegraphics{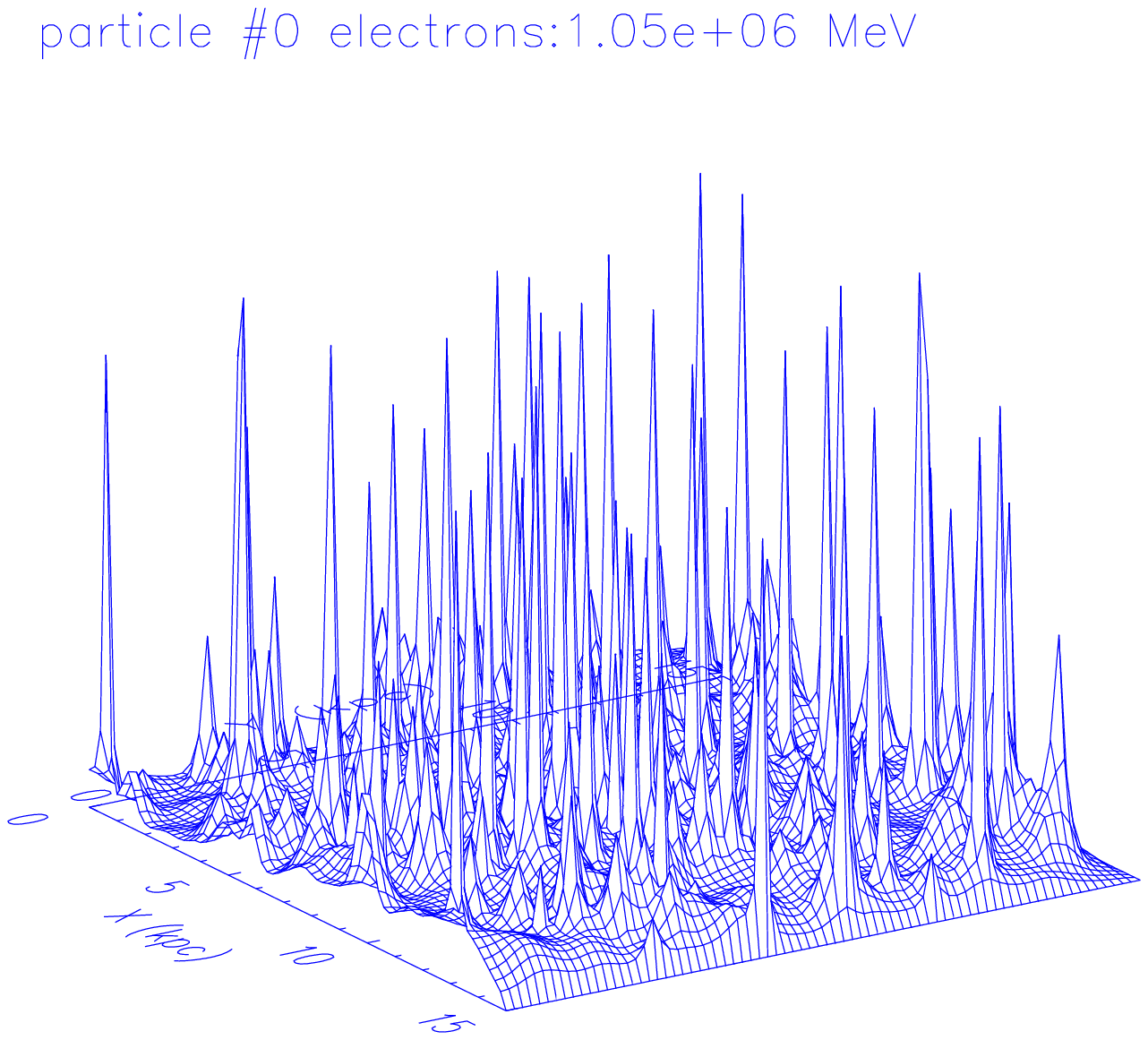} 
\caption{{\bf On the left:} The galactic plane flux of protons with E = 
64 MeV, 8.2GeV and 1 TeV, as indicated above the respective picture. At 
high energies the assumed supernova remnant sources are pronounced, at 
low energies the particle density degradation due to particle energy 
losses is clearly visible. {\bf On the right:} The galactic plane flux 
of electrons with E = 1 GeV, 16 GeV and 1 TeV. (figures from Strong \& 
Moskalenko p.1942, p.1964)} 
\end{figure*} 
 
A new perspective for investigations of the particle fragmentation 
processes within the galactic disk is an extensive numerical 
GALPROP code developed by Mos\-ka\-len\-ko, Strong and collaborators (OG 
1.3-p.1836, p.1868, p.1942, p.1964). It solves the cosmic ray 
propagation equations on a spatial grid. In principle it allows one to 
consider several realistic factors in the modelling (OG 1.3-p.1942), 
including -- in its 2D or 3D version -- a non-uniform magnetic field and 
an ambient gas structure, a wealth of particle interactions with the 
respective cross-sections, and/or realistic distribution of cosmic ray 
sources, as illustrated in figure 2. The proposed approach enables 
investigation of several characteristics of cosmic ray propagation and 
-- at present -- it is the best available tool to study cosmic ray 
abundances and ages. Of course, the approach is subject to the same 
problems with insufficient knowledge of interaction cross-sections, but 
it is able to check a substantial number of interaction channels and it 
can test the role of other factors influencing the cosmic ray 
propagation. 
 
A few papers discussed constraints on the galactic propagation models 
based on antiproton measurements (Moskalenko et al. p.1868, Sina et al. 
p.1873 and Molnar \& Simon p.1877). Difficulties met with the fits may 
indicate need for improving the propagation models considered.

\subsection{Propagation models and parameters} 
 
A number of interesting papers on physics of cosmic ray propagation 
discussed issues like diffusive propagation models in the Galaxy (OG 
1.3-p.1908, p.1926, p.1933, p.1921, p.1873, p.1885), some aspects of 
particle interaction with the turbulent medium (OG 1.3-p.1920, p.1889, 
p.1896, p.1900, p.1903, p.1904, p.1852), parameters of the cosmic ray 
flux from a local supernova remnant (OG 1.3-p.1807, p.1810), parameters 
of galactic cosmic ray propagation (OG 1.3-p.1947, p.1819, p.1923, 
p.1856) and information about recent developments in the above mentioned 
extensive numerical code to treat galactic cosmic ray propagation (OG 
1.3-p.1942). 
 
An increasing amount of accurate data on cosmic ray abundances and 
improving knowledge of the interstellar medium and the magnetic field in 
the galactic disk limit a role of exact solutions of the diffusion or 
diffusion-convection equation. Nevertheless such solutions will always 
play an important role in qualitative understanding of particle 
transport and will provide good tests for checking accuracy of numerical 
codes. A set of such solutions for galactic cosmic ray transport was 
presented by Shibata (p.1908, p.1926, p.1933) and Buesching et al. 
(p.1921). A new approach to the cosmic ray transport considering the 
anomalous diffusion was studied in numerous papers of Lagutin and 
collaborators (p.1852, p.1884, p.1889, p.1896, p.1900, p.1920). 
Unfortunately, the relation of the proposed analytic approach to the real 
conditions is at least unclear. The progress in the near future is 
expected to be based on numerical studies (like still developing 
GALPROP - see OG 1.3-p.1942 and Fig. 2) including a growing amount of 
data on the interstellar medium and new, more precise information 
about nuclear cross-sections.

\subsection{Wide range spectrum} 
 
Several papers discussed features observed in the wide energy range 
spectrum, concentrating on the so called `knee' and `ankle' features (OG 
1.2 - p.1804, OG 1.3-p.1979, p.1995, p.1968), propagation of UHE cosmic 
rays and/or their possible galactic origin (OG 1.3-p.1972, p.1976, 
p.1999, p.1991, p.1951), finally on energy losses of propagating particles 
(OG 1.3-1951), in particular iron nuclei (OG 1.3-p.1987).

Studies of the wide range energy spectrum for the heavy cosmic ray 
component have concentrated for years on the features seen in the spectrum, 
in particular the knee at $10^{15}$ - $10^{16}$ eV and the ankle above 
$10^{18}$ eV, and the power law sections between them. A growing amount 
of KASKADE data seems to confirm (Schatz HE 1.02-p.76, Fig.~3) earlier 
considerations of Erlykin and Wolfendale (cf. OG 1.2-p.1804). Based on 
the inhomogeneous data then available they noted a relatively sharp 
change of the spectrum inclination at the knee, incompatible with the 
hypothesis of distributed galactic sources with varying upper energy 
cut-off and the diffusive transport averaging these particle 
populations. Additionally, discrete wavy features noted by these authors 
at the knee were also difficult to understand within the above model. As 
a simple solution for both these observational problems Erlykin \& 
Wolfendale propose that our local cosmic ray spectrum is an averaged 
galactic spectrum modified by a single local source (a supernova 
remnant). During the present conference they presented an extended 
discussion of several aspects of their improved model (OG 1.2-p.1804), 
including particle anisotropy and temporal variations for a local 
supernova remnant (OG 1.3-p.1807, p.1810) and its electron component (OG 
1.2-p.1812). The present growing evidence of reality of the mentioned 
spectral features makes their study an important one, opening a new 
phenomenology for the cosmic ray data. 
 
\begin{figure}                       
\vspace{8.0cm} 
\includegraphics{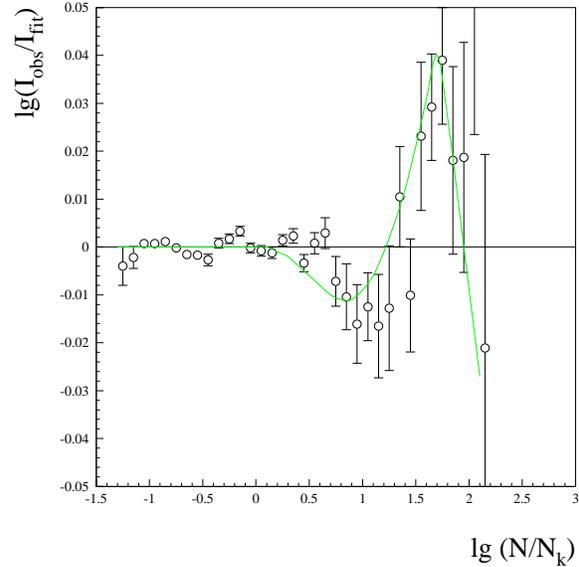} 
\caption{Differences between the KASKADE data and 5 parameter knee fit values 
(see Schatz HE 1.02-76 for a detailed description).} 
\end{figure}

The question of primary UHE particles generating atmospheric showers in 
the energy range $\sim 10^{20}$ eV is still under debate. An interesting 
contribution to this discussion is presented by Wibig \& Wolfendale 
(p.1987), who investigate propagation characteristics of ultra high 
energy heavy nuclei. They note that such nuclei interacting with the 
microwave background photons lose energy very slowly, mostly by gradual 
stripping of single nucleons from the original nucleus. In their 
modelling for a uniform source distribution in space and injection 
of high energy particles with 50\% Fe and 50\% O, the observed 
distribution at Earth extends to higher energies than the one for a pure 
proton injection (Fig.4). A growing number of studies pointing at pulsar 
sources of $10^{20}$ eV iron nuclei (see below) makes this study of 
particular interest.

\begin{figure}                     
\vspace{88mm} 
\includegraphics{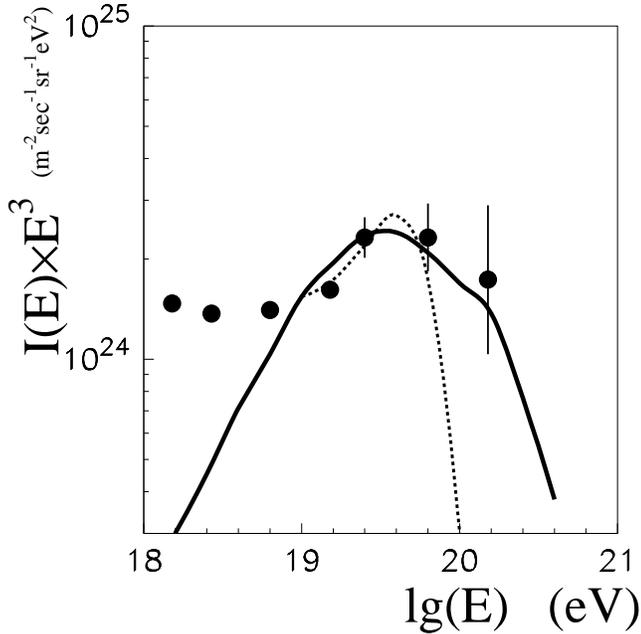} 
\caption{The measured UHECR energy spectrum (points) compared to 
predictions for a uniform source distribution and the initial 
composition of 50\% O and 50\% Fe (full line). Dashed curve shows GZK 
effect obtained for a universal proton flux. (from Wibig \& Wolfendale 
p.1987)} 
\end{figure}

\subsection{Cosmic ray electrons and $\gamma$-rays} 
 
For cosmic ray electrons their possible relation to the Gould's belt 
young stellar population was discussed in OG 1.3-p.1912 and a local 
supernova source in OG 1.2-p.1812 (see also OG 1.3-p.1964), numerical 
modelling of the electron injection and acceleration was presented in OG 
1.3-p.1848 and its relation to the galactic magnetic field structure via 
radio astronomical observations was studied in OG 1.3-p.1924. Processes 
of $\gamma$-ray radiation in the Galaxy were numerically modelled in OG 
1.3-p.1964. 
 
Growing evidence of a possible local electron component from the Gould's 
belt of recent star formation (Pohl et al. p.1912) or a local supernova 
remnant (Erlykin \& Wolfendale p.1812) becomes an option which enables 
one to explain the excess $\gamma$-ray emission above 1 GeV as a result 
of inverse-Compton radiation of the electron component with a spatially 
varying spectral index. Although inconclusive, the studies support the 
possibility that the cosmic rays observed at Earth involve two 
components - the smooth large scale galactic distribution and the more 
or less local component due to sources situated nearby. To study such 
non-uniform electron distribution Strong \& Moskalenko used numerical 
modelling (cf. Fig. 2). Due to radiation losses the obtained electron 
distribution can be more inhomogeneous than the one for protons of the 
same energy.

\subsection{Interactions and cross-sections} 
 
An important issue for modelling the cosmic ray propagation is 
derivation (measurements !) of interaction cross-sections. 
Unfortunately, this important problem was discussed in only a few papers 
(OG 1.3-p.1955, p.\-1956, p.1960), and does not represent the full range 
of the work done in this field.

\section{Cosmic ray acceleration} 
 
Continued progress in studying the cosmic ray acceleration processes was 
represented in Hamburg by a number of interesting contributions. A 
general impression one could draw from these presentations is that in 
the `classic' shock acceleration models, both non-relativistic and 
relativistic ones, there is still space for including effects which can 
substantially modify the acceleration process. We also note that only a 
few authors decided to study old, extremely important but difficult and 
time consuming problems of particle injection and the collisionless 
shock structure at  the microscopic plasma level.

\subsection{Electron injection} 
 
Two research groups discussed this issue for quasi-per\-pendicular 
shocks using similar {\it particle in cell} (PIC) simulations 
(Schmitz et al. p.2022, Drury et al. p.2096). The acceleration 
process is caused by ions reflected from the shock, which due to a two 
stream instability generate electrostatic fluctuations. Complicated 
interaction of electrons with these waves leads to their acceleration, 
generating a power law distribution up to velocities much exceeding the 
initial ions velocity. Schmitz et al. investigated the influence of the 
plasma beta on the acceleration efficiency and found that the low beta 
shocks were more efficient in accelerating electrons (Fig.~5). In my 
opinion the above approach, continuing a few earlier papers intended to 
understand the microphysics of the acceleration process, provides a 
promising attempt to reach real progress in the study of particle shock 
acceleration, surpassing relatively primitive test particle studies 
since the discovery of the diffusive shock acceleration process.

\begin{figure}                      
\vspace{62mm} 
\includegraphics{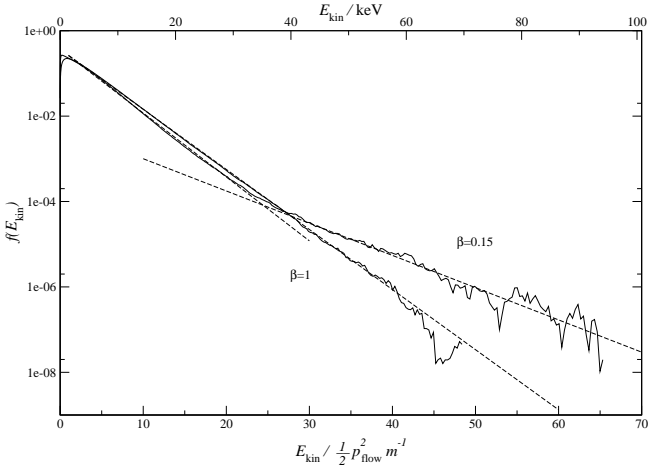} 
\caption{The energy distribution of accelerated electrons for $\beta = 
0.15$ and $1.0$ in simulations of Schmitz et al. (p.2022).} 
\end{figure}

\subsection{Non-relativistic shock acceleration} 
 
Among various aspects of non-relativistic shock acceleration a number of 
presentations were based on hydrodynamical acceleration models (OG 1.4 - 
p.2010, p.2018, p.2046, p.2050, p.2088), modelling of particle-wave 
interactions at the shock wave (OG 1.4-p.2054), anomalous cosmic ray 
transport in perpendicular shocks (OG 1.4-p.2066) and numerical study of 
the ordinary diffusive shock acceleration (OG 1.4-p.2084). Let me note a 
few interesting results. 
 
Malkov et al. (p.2010) considered non-linear particle acceleration at 
a supernova remnant shock wave. They studied the combined effect of 
acceleration non-linearity leading to the shock profile modification and 
propagation of Alfv\'en waves that are responsible for particle 
confinement near the shock. In this process refraction effects in the 
upstream pre-compression of the flow lead to decreasing wave amplitude 
in the resonance range near the spectrum cut-off. As a result the upper 
energy limit of accelerated particles is also decreased (cf. Fig 6). The 
discussed new effect provides an additional constraint on attempts to 
explain cosmic rays up to the knee as produced by SNRs.

\begin{figure}                      
\vspace{62mm} 
\includegraphics{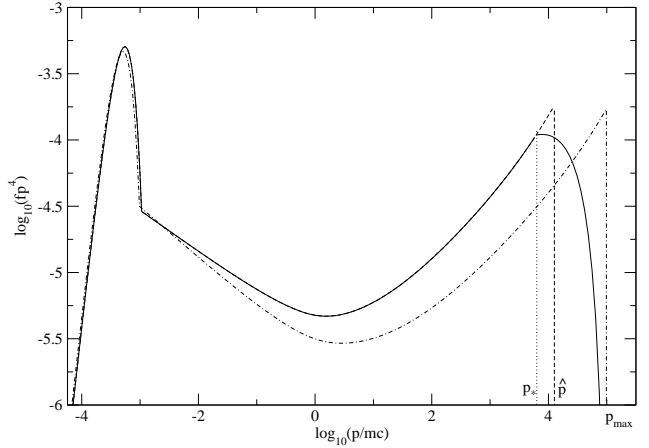} 
\caption{Non-linear spectra considered by Malkov et al. (p.2010). The 
dash-dotted line represents an analytic solution for a strong $M = 150$ 
shock and $p_{max} = 10^5 mc$. The spectrum drawn with the full line 
demonstrates the effect of wave compression on this spectrum, decreasing 
the energy cut-off to $p_*$. A dashed line represents an analytic 
solution with $p_{max} = \hat{p}$, which matches the modified spectrum.} 
\end{figure}

Another interesting study of wave propagation near the shock was 
presented by Vainio (p.2054), who considered simple unmodified shocks, 
but derived exactly wave propagation under the shock conditions. The 
upstream cosmic ray generated waves, which move forward, in the same 
direction as the shock (Vainio describes these waves as having 
a cross-helicity = 1), generate both forward and backward waves when 
transmitted downstream of the shock. A finite particle pressure 
downstream of the shock can modify the wave forward-backward asymmetry 
to the situation with waves propagating mostly backward. As a result the 
acceleration region with the large scattering centre compression ratio 
has finite width behind the shock, which regulates the acceleration 
efficiency. The computations are performed for parallel shocks, and are 
thus of limited validity, but point out a new factor influencing the 
acceleration process at small and medium Mach number shocks. 
 
The role of realistic, oblique magnetic field configuration at the 
non-linear shock acceleration was studied by Zakharian et al. (p.2018), 
who considered a spherical supernova remnant shock wave expanding in the 
uniform ambient magnetic field. They solve ideal magnetohydrodynamic 
equations with the cosmic ray force term included and the kinetic 
equation for the cosmic ray distribution, including anisotropic 
diffusion and drifts in the magnetic field near the shock. The obtained 
spectra depend to a large degree on the local magnetic field structure, 
with a much larger cut-off energy near the `equator' with the 
quasi-perpendicular shock configuration. In their `simple' model the 
particle cut-off energy can differ along the shock by a factor $> 10^3$, 
as presented at Fig.~7. It is worth to note, however, that for large 
amplitude turbulence near the shock this difference can be substantially 
smaller.

\begin{figure}                                
\vspace{90mm} 
\includegraphics{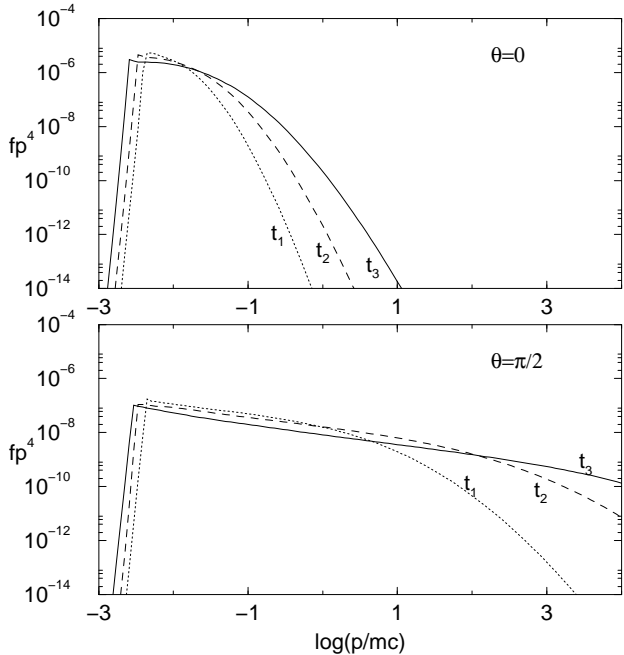} 
\caption{The cosmic ray energy distributions of particles accelerated 
at the quasi-parallel ($\theta = 0$) and the quasi-perpendicular shock 
($\theta = \pi /2$) region for the uniform ratio of the diffusion 
coefficients $\kappa_\perp / \kappa_\| = 0.2$. Results for three times 
$t_1 < t_2 < t_3$ are presented (from Zakharian et al. p.2018).} 
\end{figure} 
 
Particle Fermi acceleration at perpendicular shocks depends to a large 
degree on particle ability to diffuse across the mean magnetic field, 
and, thus, on magnetic field line wandering in a turbulent medium near the 
shock. Ragot (p.2066) applied her computations of line wandering for 
different spectra of MHD turbulence, showing an anomalous diffusion 
behaviour depending on the considered spectrum, to evaluate the particle 
acceleration time scales for shocks with oblique magnetic fields. 
Depending on the shock parameters, she found the possibility of large 
departures of the obtained time scales from the values derived in a 
traditional way. The considered processes of anomalous transport enable 
to explain variations of the radio flux from the supernova remnant 
SN1987A.

\subsection{Relativistic shock acceleration} 
 
Several authors attempted to use Monte Carlo methods to study the 
acceleration process within the test particle approach (OG 1.4-p.2006, 
p.2014, 2026, OG 2.4-p.2742, cf. a comment in Appendix) and including 
gas hydrodynamics (OG 1.4-p.2058), an acceleration `box model' was 
extended to relativistic shocks (OG 1.4-p.2043), a general discussion is 
included in OG 1.4-p.2039, finally, an attempt to discuss microphysics 
of the acceleration process was presented in OG 2.3-p.2713. 
 
The microphysics of the relativistic shock acceleration is an issue 
appearing to be essential for any reasonable study of energetic particle 
generation at such shocks. At the present meeting Pohl et al. (p.2713) 
studied electrostatic instability at a parallel relativistic shock leading 
to particle acceleration and isotropization. Their work performed with 
simple analytic means allows one to note the possibly complicated nature 
of the acceleration processes at such shocks. This study, accompanying a 
few other recently published papers, makes it clear that to achieve 
better understanding of the acceleration process at relativistic shocks 
one needs an immense amount of work devoted to studying the acceleration 
process collisionless microphysics. 
 
Dermer (p.2039) included in his talk a discussion of the particle 
acceleration in relativistic turbulence inevitably existing downstream 
of the shock. The second-order Fermi acceleration in such high amplitude 
turbulence can be extremely fast and efficient. The process should be 
studied in more detail to make the Dermer calculations more realistic, 
e.g. existence of long, high amplitude resonance waves important for 
acceleration of UHE CR particles can be questionable for relativistic 
shocks, as the shock generated waves are expected to be dominated by the 
short wave component.

\subsection{Other cosmic ray acceleration scenarios} 
 
Acceleration processes assumed to act in astrophysical objects were also 
discussed for jets (OG 1.4-p.2030, p.2034, OG 2.3-p.2701), pulsars (OG 
1.4-p.2105, p.2092), active galactic nuclei (OG 1.4 - p.2100), \, 
supergalactic structures (OG 1.4-p.2062), dust acceleration for 
explaining the observed $\gamma$-ray line width (OG 1.4-p.2077). More 
general discussion was presented in OG 1.4-p.2070, p. 2074, a role of 
new physics with the Lorentz symmetry violation was considered in OG 
1.4-p.2103, p.2104 . 
 
Giller \& Lipski (p.2092) investigated a simple model of pulsar 
acceleration to model the galactic cosmic ray distribution at energies above 
$\sim 10^{15}$ eV. They assumed that the pulsar rotational energy is 
fully transferred into energetic particles. These particles are injected 
into the interstellar medium at the highest possible energy at a given phase 
of pulsar evolution (i.e. the energy scale decreases when the pulsar 
rotation slows down and its magnetic field decays). Thus the observed 
spectrum should be an average over the pulsar population. The 
authors show that the data on cosmic rays and pulsars within our Galaxy 
are consistent with this model. 
 
An interesting study of observational effects related to dust grain 
shock acceleration measured by gamma-ray line shape was presented by 
Kretschmer et al. (p.2077). The observed $^{26}$Al $1.809$ MeV line 
width of $\approx 5$ keV is explained by kinematic velocities of the 
bulk of aluminium atoms in the studied region containing young stars 
with powerful stellar winds and supernova remnant shock waves. The 
authors consider acceleration of dust grains to large velocities as 
parts of such flows, their destruction into individual nuclei after 
transmission downstream of the respective shock, and the velocity evolution 
of these nuclei afterwards. A long life-time of fast nuclei can explain 
the observed line widths. If the shock acceleration generates 
a population of even more energetic nuclei, a `blue' wing should 
accompany the thermal line, as can be tested in future observations.

\section{Final remarks} 
 
Let me express a few impressions from my present work. The first is 
that it provided only a partial description of achievements on the 
propagation and acceleration issues at the 27th ICRC because a 
substantial part of the work devoted to these problems was presented 
during the conference SH sessions. It seems to me that the present 
division, reflected partly in separate sets of journals, where the 
respective papers are published (geophysical ones for the heliospheric 
studies, and astrophysical ones for investigation of a more distant 
space), is an obstacle for common research effort in the considered 
field. 
 
As clearly seen at this conference further progress in modelling 
cosmic ray galactic propagation requires substantial improvement of our 
knowledge of interaction cross sections, the magnetic field structure in 
the Galaxy and local effects. Until such progress occurs, of particular 
interest would be simultaneous study of propagation for large number of 
different particles (various isotopes, antiprotons and electrons or 
positrons) to get a larger number of constraints for parameters of the 
considered model.

In study of acceleration processes progress is to be expected from the 
detailed studies of `microphysics' of the acceleration process, 
considering details of collisionless shocks and MHD turbulence. In 
particular the processes of relativistic shock acceleration are far from 
being fully understood. However, even for the relatively well studied 
domain of non-relativistic shocks new processes were considered 
at this meeting showing sometimes qualitatively new possibilities.

\begin{acknowledgements} 
I am grateful to all authors, who provided figures presented in the 
present report. This work was supported by the {\it Komitet Bada\'n 
Naukowych} through the grant PB~258/P03/99/17~. 
\end{acknowledgements} 
 
\appendix 
 
\section{A comment on Monte Carlo modelling of particle Fermi 
acceleration at relativistic shocks} 
 
Let us recall why the Monte Carlo modelling of particle acceleration at 
relativistic shocks applying the method of {\it discrete isotropic (or large 
angle)} scattering of particle momentum leads to unphysical results. The 
first order Fermi acceleration process involves particles wandering near 
the shock front and crossing it occasionally upstream or downstream. A 
particle transmitted upstream of the shock starts to move in front of 
it with the velocity component along the shock normal, $v_\|$, 
larger than the shock velocity with respect to the upstream medium, 
 
$$u = c \sqrt{1-{1 \over \gamma^2}} \approx c \left( 1 - {1 \over 2 
\gamma^2} \right)  \quad , \eqno(A1)$$ 
 
\noindent 
where $\gamma$ is the upstream shock Lorentz factor. Here we 
consider relativistic shocks with $\gamma \gg 1$. After the upstream 
perturbed magnetic field changes the particle momentum through a small angle 
$\theta > 1/\gamma$ with respect to the shock normal the shock starts to 
approach the particle to overtake it quickly, usually for $\theta < 
2/\gamma$. While upstream of the shock the particle distance from the 
shock is on average not larger than $\sim | v_\| - u | \Delta t < 
r_g/\gamma^3$, where $r_g$ is the particle gyroradius and the 
particle upstream residence time $\Delta t \sim 2/\gamma \cdot (r_g/c)$, 
for the magnetic field not strictly uniform and parallel to the shock 
normal. Due to the high anisotropy of particles interacting with the shock 
an average energy gain per one cycle {\it downstream - upstream - 
downstream} is comparable to the particle original energy, $\Delta E 
\sim E$ . 
 
The above energy gain is much smaller than the highest possible effect 
$\Delta E / E \sim \gamma^2$ occurring when particle momenta, after 
crossing the shock, are instantaneously scattered on large angles. One 
should note that in order to change the upstream particle momentum back, 
toward the shock, until the shock overtakes it, the particle has to move 
the distance $> r_g$ off the shock. It requires that the particle moves 
along the shock normal at least for a time $2 (r_g/c) \gamma^2$, which 
is a factor of $\gamma^3$ larger than the mean residence time in 
realistic magnetic field. For large $\gamma$ the probability of having 
such conditions, leading to the mean energy gain in the Fermi process 
$\Delta E / E \sim \gamma^2$, is negligible.

\end{document}